\input{aipcheck}

\documentclass[
    ,final            
  ]
  {aipproc}

\layoutstyle{8x11single}
\newcommand{\be}{\begin{equation}}
\newcommand{\ee}{\end{equation}}
\newcommand{\bea}{\begin{eqnarray}}
\newcommand{\eea}{\end{eqnarray}}
\newcommand{\bref}[1]{(\ref{#1})}

\begin{document}

\title{No-go theorems and GUT \footnote{Opening address given at GUT2012}}

\classification{12.60.-i, 12.10.-g, 11.10.Kk}
\keywords      {<GUT, New Physics, >}

\author{Takeshi FUKUYAMA}{
  address={Department of Physics and R-GIRO, Ritsumeikan University,\\
 Kusatsu, Shiga,
525-8577, Japan}
}

\begin{abstract}
 We very briefly discuss the role of no-go theorems in constructing models of
new physics beyond the SM.
\end{abstract}

\maketitle


\section{Opening Address}

  Good Morning everyone. We heartily welcome all participants.
As you well know, we suffered from the Tohoku Earthquake just one year before,
and lost twenty thousand precious lives. More than three hundred thousands peoples are still forced to live as refugees. We all share their pain and all will recover together with them as after World War II.

We received many warm letters and offer of help from abroad at that time, which encouraged us very strongly. We express our sincere thanks here to every one.\\

This workshop is the second one succeeding from GUT07 held at
Ritsumeikan Univ. on December 15-17 2007.
More than four years have passed since that time and many new facts were found during this time. There are signals of new physics beyond the Standard Model (SM): muon anomalous magnetic moments, top quark forward-backward asymmetry, $U_{bs}$ etc.
The presences of Dark Matter and Dark Energy have also been established.
The recent LHC results have also had a great impacts.
\section{No-Go Theorem in Model Building}
I have very short time and will comment on the prospect of GUT restricting myself only to the topic of the implications of No-Go theorems in GUT model building
(Fig.1).
The vertical line indicates the structure of quark-leptons of one generation,
the horizontal line is for family one.

\begin{itemize}
\item
No-go (1): Coleman-Mandula \cite{C-M} theorem with Haag-Lopuzanski-Sohnius extension \cite{H-L-S}:
The latter goes over the former by considering a rank one spinor charge,
leading to supersymmetry.
\item
No-go (2): String theory does not allow for a high dimensional Higgs \cite{Dienes}, which is connected with Heterotic string model and their perturbations.
Non perturbative F theory or composite Higgs are out of this constraint.
\item
No-go (3): (a) U(1) symmetry is necessary for SUSY breaking. (b) If there is U(1) symmetry and it is spontaneously broken, then SUSY is automatically broken \cite{Nelson}.
Reflecting these results Fallbacher et al. \cite{Ratz} concluded that no MSSM model with either a {\bf Z}$^R_{M\geq 3}$ or U(1) symmetry can be completed by a four dimensional GUT in the ultraviolet. This is a strong motivation to go to extra dimensions as a conceptual problem \cite{Fuku2}.
\item
No-go (4): This is concerned with no-go in family symmetry \cite{Koide}. This is less known unlike the other no-go theorems and is not established as no-go yet. Then we will explain it in the following
\end{itemize}
This theorem asserts that
if one Higgs contributes in each quark sector and if the Yukawa coupling has a family symmetry, then we can not bring $3\times 3$ mixing in CKM matrix.
Let me explain in more detail. Under the above conditions, the Yukawa coupling becomes 
\be
W_Y=Y^{(u)}_{ij}\overline{Q}_iH_uu_j+Y^{(d)}_{ij}\overline{Q}_iH_dd_j,
\label{Yukawa}
\ee
where $Q$ is left-handed doublet, and $u,~d$ are right-handed singlets.
Then the family symmetry,
\be
Q\rightarrow Q'=T_LQ,~u\rightarrow u'=T_R^uu,~d\rightarrow d'=T_R^dd.
\ee
gives a constraint on $Y^u$ and $Y^d$
\be
T_L^\dagger Y^uT_R^u=Y^u,~~T_L^\dagger Y^d T_R^d=Y^d.
\label{F1}
\ee
Eq.\bref{F1} leads us finally to \cite{Koide}
\begin{equation}
\left(e^{i\delta_i^u}-e^{i\delta_j^d}\right)V_{ij}=0.
\label{flavour}
\end{equation}
Here the SU(3) family transfomation is factorized as
\begin{equation}
P_f=diag\left(e^{i\delta_1^f}, ~e^{i\delta_2^f},~e^{i\delta_3^f} \right)
\end{equation}
and
\begin{equation}
T_L=U_L^uP_u\left(U_L^u\right)^\dagger=U_L^dP_d\left(U_L^d\right)^\dagger.
\end{equation}
Eq. \ref{flavour} shows that we can get only a two-family mixing, showing the inconsistency with observations.

However, a model of many-Higgs in each sector seemed to evade this no-go theorem.

In the framework of GUT, we have many Higgs doblets. For instance, the renormalizable minimal SO(10) GUT with ${\bf 10}$ and $\overline{{\bf 126}}$ Higgs field
have 4 Higgs doublets each in up-sector and down-sector \cite{F-O},
\be
H_u^{10} \equiv H_{\bf(1,2,2)}^{({\bf 1,2},\frac{1}{2})}, \,
\overline{\Delta}_u \equiv 
\overline{\Delta}_{\bf(15,2,2)}^{({\bf 1,2},\frac{1}{2})}, \, 
\Delta_u \equiv 
\Delta_{\bf(15,2,2)}^{({\bf 1,2},\frac{1}{2})}, \,
\Phi_u \equiv 
\Phi_{\bf(\overline{10},2,2)}^{({\bf 1,2},\frac{1}{2})}.
\ee
and 
\be
H_d^{10} \equiv H_{\bf(1,2,2)}^{({\bf 1,2},-\frac{1}{2})}, \,
\overline{\Delta}_d \equiv 
\overline{\Delta}_{\bf(15,2,2)}^{({\bf 1,2},-\frac{1}{2})}, \,
\Delta_d \equiv \Delta_{\bf(15,2,2)}^{({\bf 1,2},-\frac{1}{2})}, \, 
\Phi_d \equiv 
\Phi_{\bf(10,2,2)}^{({\bf 1,2},-\frac{1}{2})}.
\ee
Here $H,~\Delta,~\overline{|Delta},^\Phi$ represent the elements from ${\bf 10},~{\bf 126},~
\overline{{\bf 126}},~{\bf 210}$, resectively. Also subscript (superscript) shows Pati-Salam (the SM) element. 
We can diagonalize the mass matrix, $M_{\mathsf{doublet}}$ 
by a bi-unitary transformation.   
\be
U^{\ast} \,M_{\mathsf{doublet}} \,V^{\dagger}
= {\mathrm{diag}}(0, M_{126}, M_2, M_3).    
\ee
Then the mass eigenstates are written as  
\bea
\left(H_u, 
\, {\mathsf{h}}_u^1, \, {\mathsf{h}}_u^2, \, {\mathsf{h}}_u^3 \right)
&=& 
\left(H_u^{10}, {\overline{\Delta}}_u, \Delta_u, 
\Phi_u \right) \, U^{\mathsf{T}}, 
\nonumber\\
\left(H_d, 
\, {\mathsf{h}}_d^1, \, {\mathsf{h}}_d^2, \, {\mathsf{h}}_d^3 \right)
&=&
\left(H_d^{10}, \Delta_d, {\overline{\Delta}}_d, 
\Phi_d \right) \, V^{\mathsf{T}}. 
\label{UV}
\eea
Here $H_u,~H_d$ are MSSM light Higgs doublets. However, these Higgs fields
have no family quantum number and make any change in no-go theorem.

As far as we do not break flavour symmetry explicitly, we are forced to impose family quantum numbers to Higgs fields. However, this implies that Yukawa coupling constant is not a constant but becomes a field.

 Let us consider, for instance, the case that $SU(3)_F$ and $Q,~u,~d$ belong to ${\bf 3}$-representation. Then $SU(3)_F$ invariance requires that Higgs fields is ${\bf 3}$ or $\overline{{\bf 6}}$. However, we must construct tensor $Y_{ij}$
in terms of Higgs fields with family quantum numbers (flavons) $\phi^i$. Then the invariant Yukawa can be constructe by $\overline{{\bf 3}}$ as \cite{King}
\be
W_Y^a=\frac{1}{M_a^2}\overline{Q}_i\phi_a^iu_j\phi_a^jH_a,~~(a=u,d).
\ee
Alternatively we may consider a dimension five operator by $\overline{{\bf 6}}$ and ${\bf 3}$ Higgs
\be
\phi^i\phi^j\rightarrow \phi^{(ij)},~~\epsilon^{ijk}\phi_k.
\ee
Regardless too this, we must consider unrenormalizable terms, and gauge and family groups are still independent. Should we accept this as an intermediate developing stage ?

If we include Higgs in matter multiplets, we can attach family index to Higgs
like in some of $E_6$ models \cite{Matsuoka},
\be
{\bf 27}_i=(Q,u^c,d^c,L,\nu^c,e^c,H_u,H_d,g,g^c,S)_i.
\ee
However, family structure was considered from purely phenomenological approach.
Thus GUT and family structures are left as independent ingredients.
Lagrangian construction in nonlinear representation from higher gauge group is suggestive \cite{Kugo}.
Is family symmetry gauged but still independent on horizontal gauge symmetry \cite{Mohapatra} ?

We hoped some no-go theorem in family symmetry which led us to some conceptual relations between these two ingredients. In this sence, we must still need no-go theorem in this area. 

\begin{figure}
  \includegraphics[height=.3\textheight]{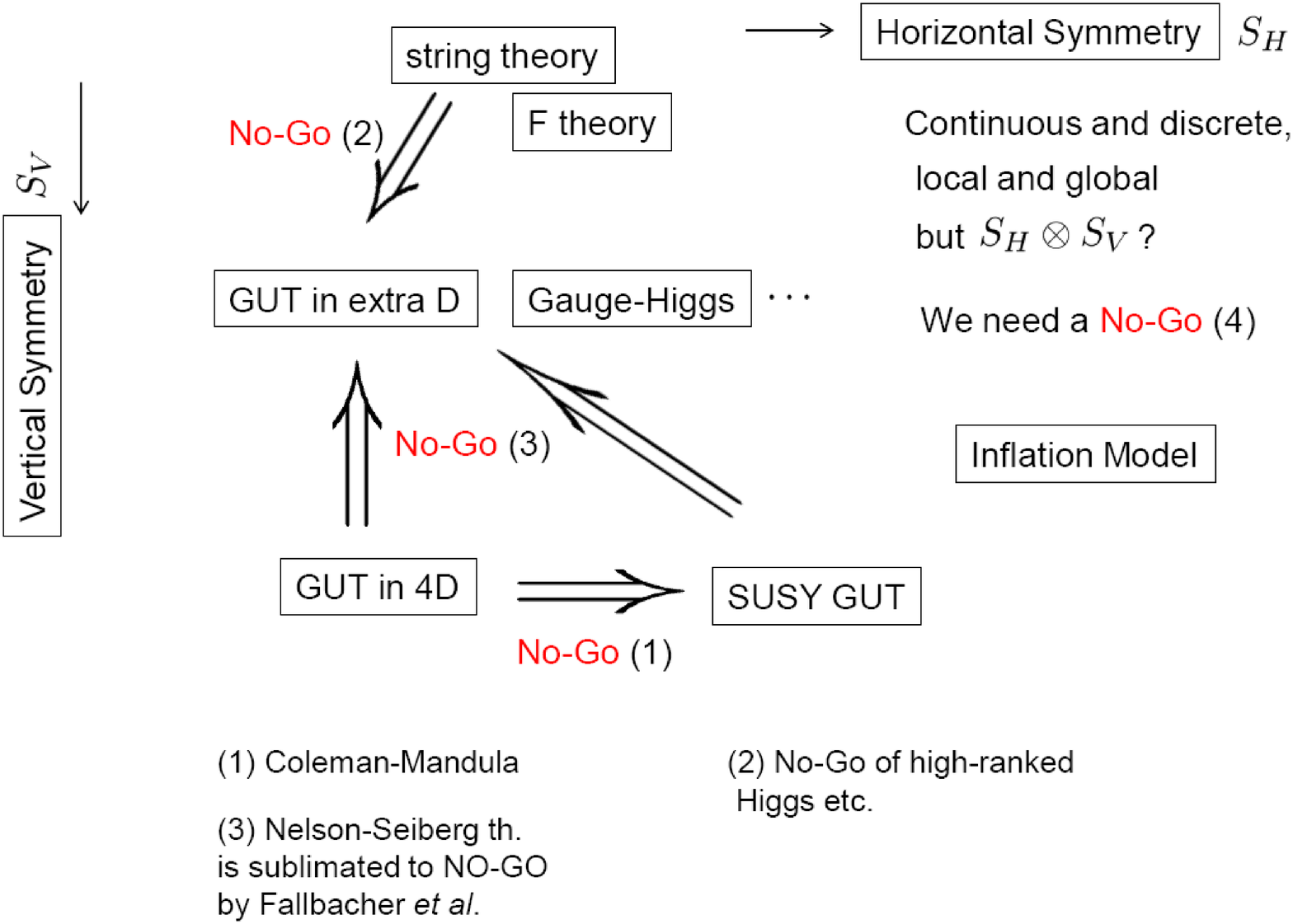}
  \caption{Schematic diagram of no-go theorems}
\end{figure}

\begin{theacknowledgments}
The author is grateful to all participants. He also thanks Koide for useful discussions. This work is partly supported by the Grant-in-Aid for Scientific Research from the Ministry of Education, Science and Culture of Japan (No.020540282 and No. 21104004). 
\end{theacknowledgments}





\IfFileExists{\jobname.bbl}{}
 {\typeout{}
  \typeout{******************************************}
  \typeout{** Please run "bibtex \jobname" to optain}
  \typeout{** the bibliography and then re-run LaTeX}
  \typeout{** twice to fix the references!}
  \typeout{******************************************}
  \typeout{}
 }



\end{document}